\documentclass{sf2a-conf}
\usepackage{graphicx}

\begin{document}
\TitreGlobal{SF2A 2007}

\title{Galaxy Evolution and Star Formation Efficiency in the Last Half of the Universe}

\author{F. Combes} \address{Observatoire de Paris, LERMA (CNRS:UMR8112), 61 Av. de l'Observatoire, 75014 Paris, France} 
\author{S. Garc\'{\i}a-Burillo} \address{Observatorio Astron\'omico Nacional (OAN)-Observatorio de Madrid,
Alfonso XII, 3, 28014-Madrid, Spain}
\author{J. Braine} \address{Observatoire de Bordeaux, Universit\'e Bordeaux~I, BP 89, 33270 Floirac, France}
\author{E. Schinnerer} \address{Max-Planck-Institut f\"ur Astronomie (MPIA), K\"onigstuhl 17, 69117 Heidelberg, Germany}
\author{F. Walter$^4$} 
\author{L. Colina} \address{IEM, Consejo Superior de Investigaciones Cientificas (CSIC), Serrano 121, 28006 Madrid, Spain}
\author{M. Gerin} \address{Radioastronomie ENS, 24 Rue Lhomond, 75005 Paris, France}

\runningtitle{Star Formation Efficiency}
\setcounter{page}{1}

\index{Combes F.}
\index{Garc\'{\i}a-Burillo S.} 
\index{Braine J.}
\index{Schinnerer E.}
\index{Walter F.}
\index{Colina L.}
\index{Gerin M.}

\maketitle

\begin{abstract}
We present the results of a CO(1-0) emission survey with the IRAM 30m of 30 galaxies at moderate redshift (z $\sim$ 0.2-0.6) to 
explore galaxy evolution and in particular the star formation efficiency, in the redshift range filling the gap between local and 
very high-z objects. Our detection rate is about 50\%. One of the bright objects was mapped at high resolution with the 
IRAM interferometer, and about 50\% of the total emission found in the 27 arcsec (97 kpc) single dish beam is recovered 
by the interferometer, suggesting the presence of extended emission. The FIR-to-CO luminosity ratio is enhanced,
following the increasing trend observed between local and high-z ultra-luminous starbursts.
\end{abstract}
%
\section{Introduction}

It has been established that the star formation in the universe was 
one or two orders of magnitude larger in the past, had
a maximum around $z=2$, and then again faded towards zero beyond
$z=6-8$ (e.g. Bouwens \& Illingworth 2006). 
When the universe was half of its current age, at $z=0.7- 0.8$, 
the star formation rate was ten times larger than today.
  Observations also tend to show that the star forming
objects were not the same, in the first half of the Hubble time.
  Starbursts occur first in the bigger galaxies, that 
are then quenched, while in the second half of the universe, 
star formation occurs in smaller objects,  and this phenomenon is
called downsizing.

This strong evolution of star-forming galaxies is not
really understood, although it is suspected
that galaxy interactions and mergers are certainly one
of the main factors.  The frequency of mergers with time
has been estimated, through the perturbed morphologies,
the asymmetry index and the number of galaxy pairs as
a function of redshift.
Luminous galaxies in particular, brighter than L$_*$,
show a merger fraction of 10\% at $z=1$, and  
50\% at $z=3$ (Conselice 2006).
Massive galaxies at z=3 have 4-5 major mergers 
between z=3 and 0, and most of them occur at z $>$ 1.5.

This past active star formation in the universe 
contributes largely to the cosmic infrared background (CIB).
About half of the galaxy formation and evolution escapes 
as direct starlight, the other half is re-radiated by dust.
Galaxies contributing the most to the total CIB are z$\sim$ 1 
luminous infrared galaxies, which have intermediate stellar masses
(Dole et al 2006). The CIB has about the same brightness than 
the COB (the optical background), and 
both have only 5\% of the brightness of the CMB.
 While today normal star forming galaxies dominate
the star formation budget, the LIRGs
begin to dominate beyond z=0.7, and the ULIRGs may be even earlier
(e.g. Perez-Gonzalez et al. 2005).

\bigskip

To better understand the physics responsible for this evolution, it is
essential to measure the molecular gas content of the intermediate redshift
galaxies, and determine the Star Formation Efficiency (SFE),
as a function of redshift.  The SFE is estimated by the ratio
of the far infrared luminosity (tracer of SF) to the CO emission
(tracer of the H$_2$ content).
This efficiency appears to increase with redshift, at least for
objects detected. For instance the 
Submillimeter Galaxies (SMGs), are apparently more efficiently
forming stars than ULIRGs,
with SFR of $\sim$ 700 M$_\odot$/yr, during a starbursting phase
of 40- 200 Myr (Greve et al 2005).
A possibility is that these early objects have not yet accumulated
bulges, and mergers without bulges could be more violently active.

\section{Observation of 30 ULIRGs at intermediate z}

\begin{figure}[h]  
\begin{center}
\includegraphics[angle=-90,width=13cm]{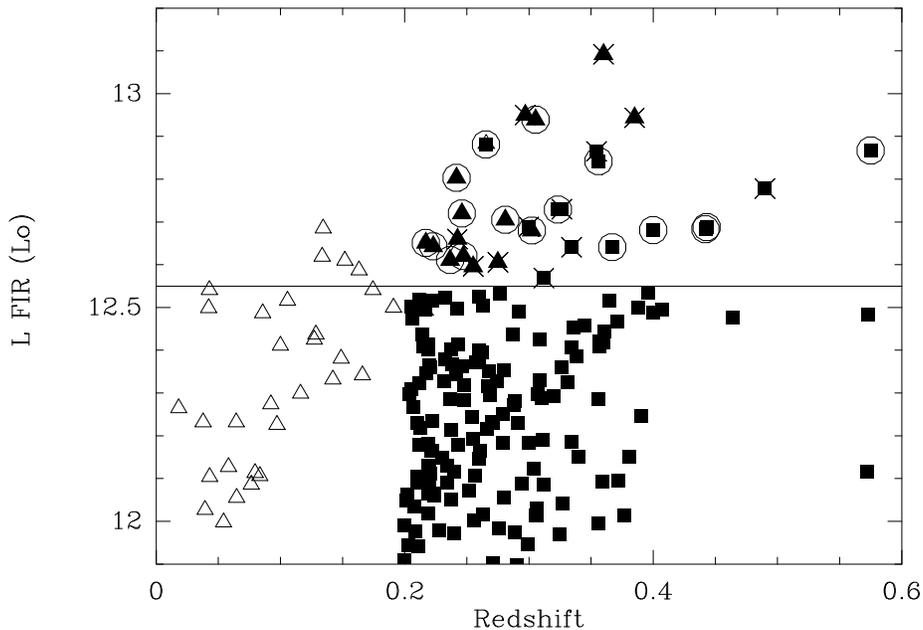}
\caption{Plot of far-infrared luminosity, $L_{\rm FIR}$, of ULIRGs as a
function of redshift, $z$, in the local sample of Solomon et al. (1997)
(open triangles) and our sample (filled symbols).  We have observed
at the IRAM-30m the points above the horizontal line (30 sources).
The circles indicate secure detections,
non-detections are marked by a cross.  }
\end{center}
\label{sample} 
\end{figure} 

Up to now, very little is known about the molecular gas content of
galaxies at moderate redshift. The local ULIRGs sample of Solomon et al (1997)
contains 37 objects, but only 2 have  redshifts of $z > 0.2$. 
We have undertaken a systematic survey of $ 0.2 < z < 0.6 $ sources,
to begin to fill the gap between low-z and high-z ($>2$) studies.
We have selected all objects (209 galaxies) at declinations greater than -12$^\circ$
 with spectroscopic redshifts and a detected 60~micron flux (from IRAS or ISO).
Most of the galaxies (and in particular the brightest ones) have
also detailed photometry in the NIR bands, from the samples by
Clements et al (1996), Kim \& Sanders (1998), Kim et al (2002)
and Stanford et al (2000).
The optical spectroscopy shows that no signs of AGN features are present
in their spectra. According to their sub-arcsec K-band images (see ref above), 
 about two-thirds of the objects are interacting galaxies.

 We observed the 30 brightest IR-luminous objects (cf Fig~1).
Among them,  17 were detected (Combes et al, in prep).
 The H$_2$ masses were derived from the CO emission,
using a conversion factor which is 5 times lower than the standard one,
as advocated by Solomon et al (1997) to be applied to ULIRGs.
 Fig~2  then compares the far infrared luminosities
with H$_2$ masses, to estimate star formation efficiencies.
The detected objects have in general high SFE with
respect to the more local sample (for which the same
conversion ratio was used). In addition to
dynamical processes being more efficient in star formation,
another possible explanation to this 
high SFE measured by $L_{\rm FIR}$/M(H$_2$) could be partly due to a non negligible 
AGN contribution to the FIR luminosity. Although we excluded in the sample 
the possible AGN (no-identification in optical spectra), there still
might be some highly embedded/obscured AGNs.
It has been established that the fraction
of AGNs in ULIRGs rapidly increases with the luminosity above
log(L/L$_\odot$)$\sim$ 12.5 (e.g. Tran et al. 2001). 

\begin{figure}[h]  
\begin{center}
\includegraphics[angle=-90,width=13cm]{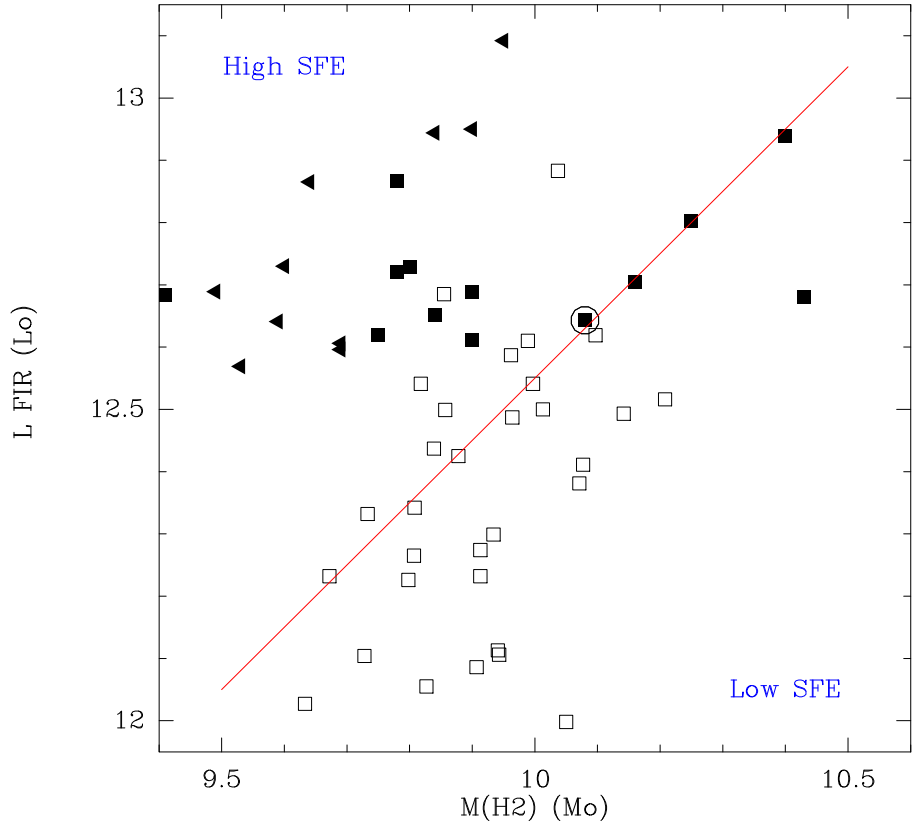}
\caption{In this plot of $L_{\rm FIR}$ versus M(H$_2$), the local sample from 
Solomon et al. (1997) is again indicated as open symbols, while our 
observed galaxies are filled symbols. Upper limits are indicated as triangles
and squares are detections. The diagram helps to identify the high star formation 
efficiency (SFE) objects in the upper left, and low SFE in the lower right. Note
the high SFE of our sample.  The galaxy observed with the IRAM interferometer
is indicated by a circle.}
\end{center} 
\label{SFE} 
\end{figure}

\section{Mapping of one of the ULIRGs at z=0.223}

We mapped with the Plateau de Bure interferometer one of
the best-detected galaxies, IRAS 11582+3020. It is an
ultra-luminous galaxy with L(IR)= 5.4 10$^{12}$ L$_\odot$, 
classified as a LINER by Kim et al. (1998). 
Rupke et al. (2005) find evidence of a superwind outflow in
this galaxy of about 15 $\rm M_{\odot}yr^{-1}$, while its SFR is
estimated at 740 $\rm M_{\odot}yr^{-1}$ from the infrared luminosity.
The red image of Kim et al. (2002) reveals some extended diffuse emission,
which could be diluted tidal tails, from a recent interaction,
while two galaxies of the same group are within 90~kpc in projection
(Fig~3). 

\begin{figure} 
\begin{minipage}[t]{.47\linewidth}
\includegraphics[width=7cm]{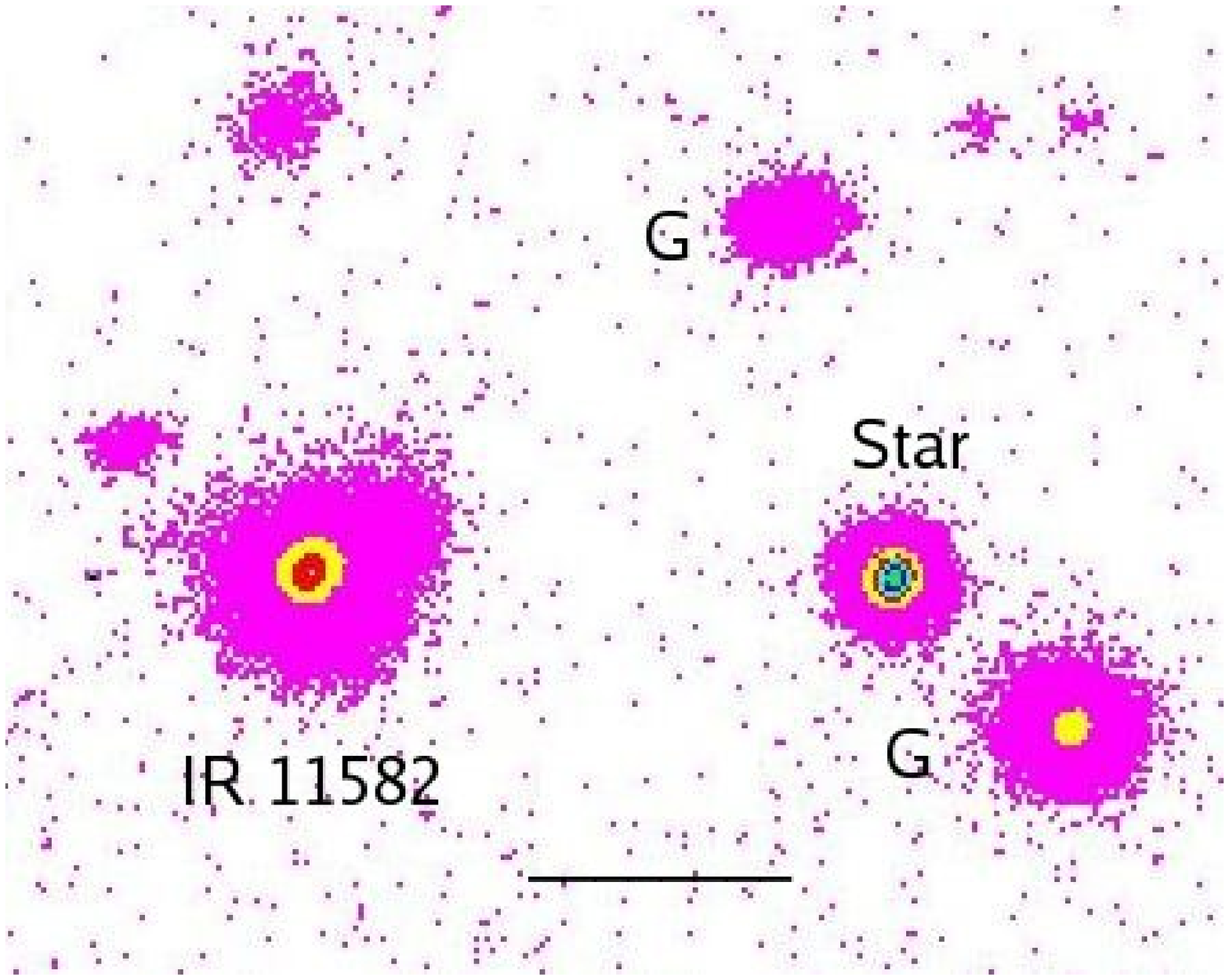}
\end{minipage}
\hfill
\begin{minipage}[t]{.47\linewidth}
\vspace{-6cm}\includegraphics[angle=-90,width=7cm]{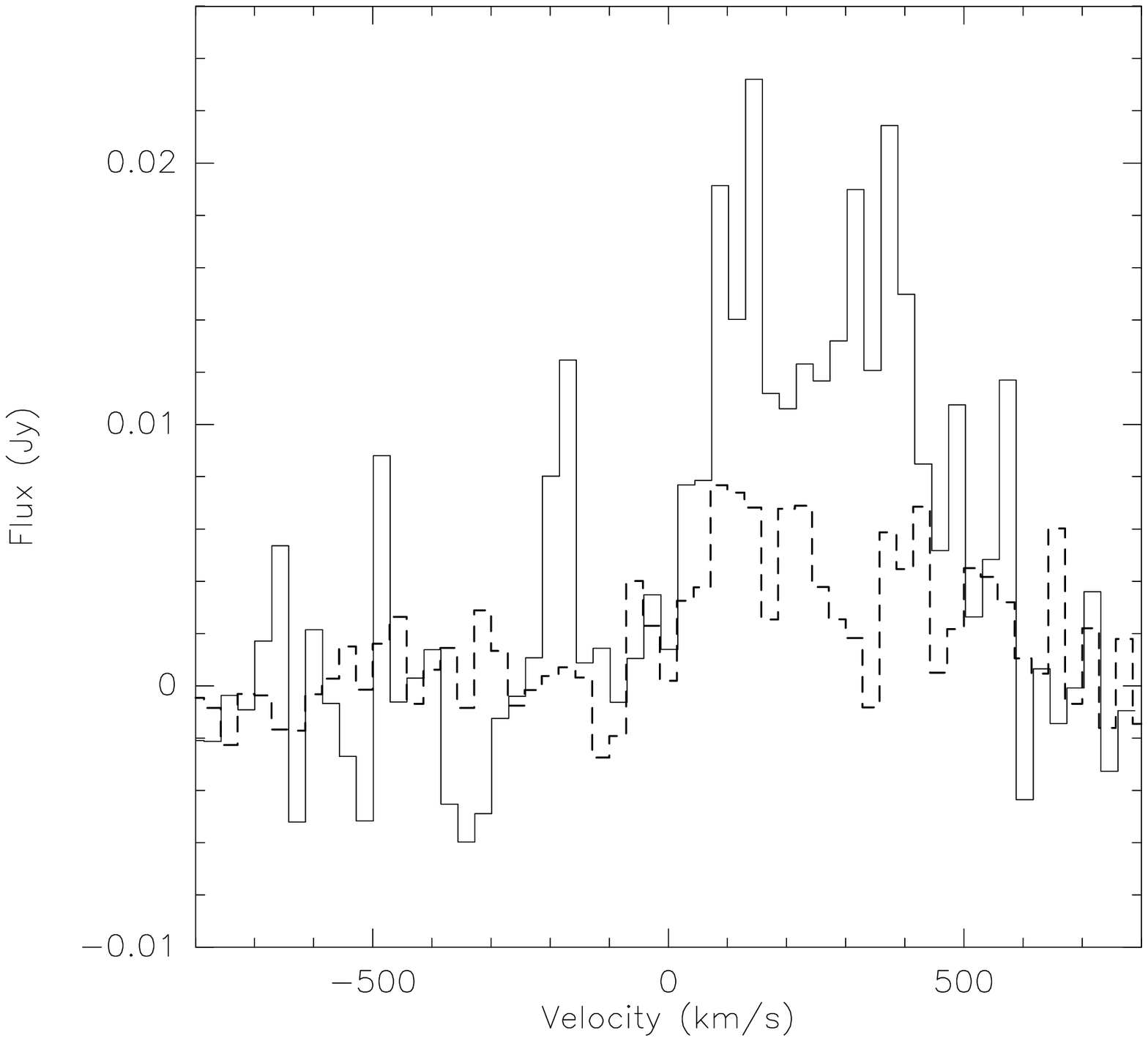} 
\end{minipage}
\caption{ {\bf Left}:  Red image of IRAS 11582+3020 from Kim et al. (2002).  The two
  objects marked "G" are galaxies in the same group (a foreground
  star is also indicated). The length of the horizontal bar at the
  bottom is 10 arcsec = 36~kpc at z=0.223.  
{\bf Right}: The IRAM 30m CO(1--0) spectrum (solid line), compared
with the integrated spectrum from the interferometer (dashed
line). The velocity is relative to z=0.223 (from Combes et al 2006).}
\label{G4} 
\end{figure}

IRAS 11582+3020 was observed in CO(1--0) first with the IRAM-30m, with 
a beam of 27" (= 97 kpc) at a frequency of  94.25 GHz, 
and then with the Plateau de Bure Interferometer (PdBI),
with a beam of  1.3''$\times$1.0''.  No continuum
emission was detected at 3mm (possible AGN) or at 1mm  (possible
dust emission), down to rms noise levels of 0.1 mJy\,beam$^{-1}$ and
0.5 mJy\,beam$^{-1}$.  Given the IRAS 100$\mu$m flux of 1.5 Jy, the upper limit 
at 1.2mm is compatible with a typical starburst SED.

The single dish and interferometer spectra are compared in Fig~3.
The PdBI flux is only $\sim$50\% of the 30m flux, suggesting
extended emission beyond 10'' (36kpc) , which is consistent with 
the extended optical structure, and 
 hints at a weak diffuse tidal tail in this perturbed system.
The H$_2$ mass derived from the 30m spectrum is
1.2  10$^{10}$ M$_\odot$, and  6 10$^9$ M$_\odot$  with the PdBI (with the low
conversion factor used for ULIRGs).

\begin{figure}[h]  
\begin{center}  
\includegraphics[width=13cm]{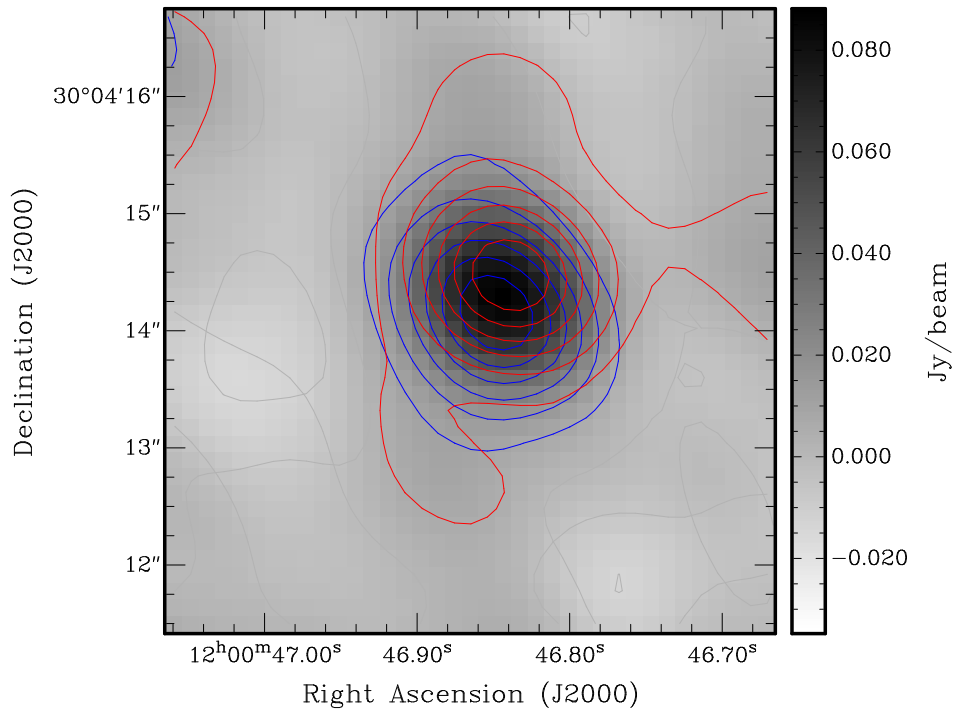}
\caption{ Blue and red velocity channels superposed
on the CO(1--0) image of IRAS 11582+3020,
obtained with the Plateau de Bure interferometer.
 A gradient of velocity is apparent.}
\end{center}  
\label{redblue} 
\end{figure}

The integrated CO(1--0) map is plotted in Fig~4, together with
the red and blue channel maps. The source is resolved at least in the
direction of the beam's minor axis, where the deconvolved size of the CO
emitting region is of the order of 0.8'', in diameter ($\sim$
3~kpc). It is possible to detect the shift of
the barycenter in each channel map, with a kinematic major axis
aligned at PA  $\sim$135$^\circ$ (Fig~5). 
Since $\sim$135$^\circ$
is also the position angle of the extended optical isophotes (Fig~3),
all available data are compatible with a post-merger relaxed system.
The estimated dynamical mass is M$_{dyn}$= 3.4 10$^{10}$ M$_\odot$.

\begin{figure}[h]  
\centering 
\includegraphics[width=13cm]{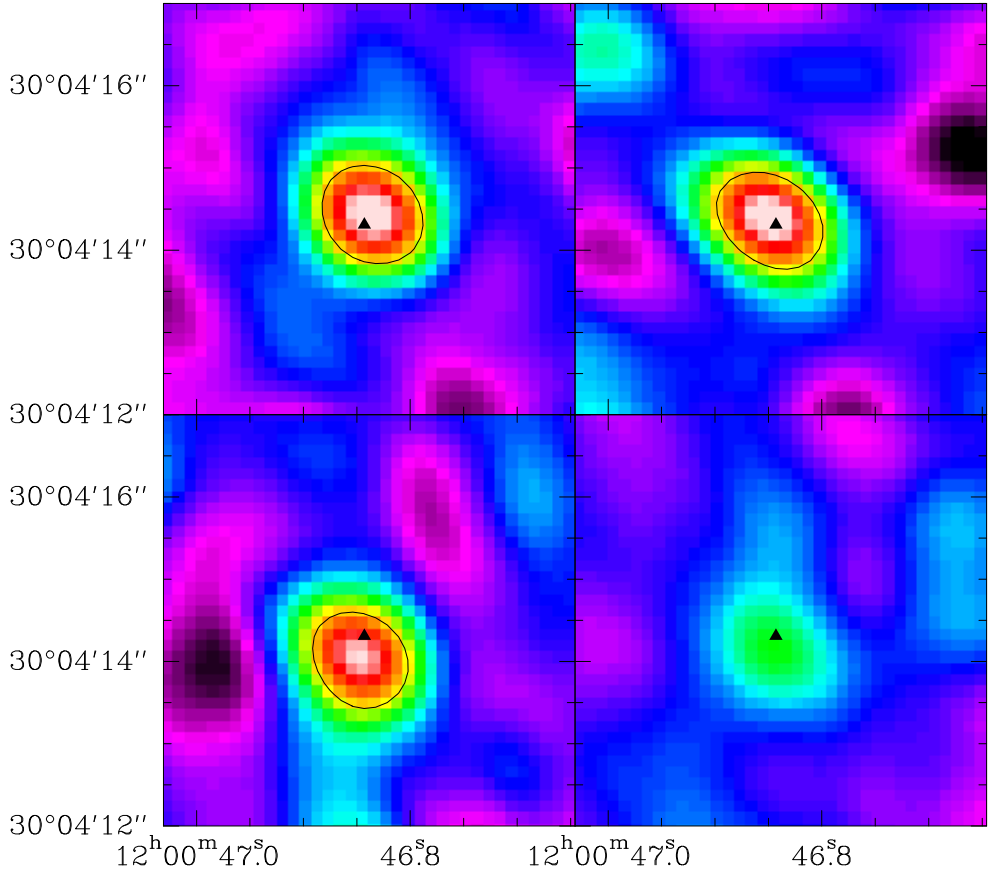}
\caption{ Four equidistant channel maps of IRAS 11582+3020,
from 500 to 0 km/s (top left to bottom right), showing the
barycenter moving along a position angle of $\sim$135$^\circ$. }
\label{chan} 
\end{figure} 

\bigskip

In summary, the CO emission in IRAS 11582+3020 
suggests the presence of two components: a nuclear disk
1.5kpc in radius with a nuclear starburst
(as usual in local starbursting galaxies, where disks of
radii= 300-800pc are found, Downes \& Solomon 1998);
and a more extended disk up to 20-30kpc diameter, which
would  correspond to the optical merger morphology.

The physical properties of IRAS 11582+3020 (size, SFE) appear to be intermediate
between those of local ULIRGs and high-z submillimeter galaxies
(SMG) as mapped by Tacconi et al. (2006). 
There might be a trend, also followed by IRAS 11582+3020, for high-z
galaxies to have a higher L(FIR)/M(H$_2$) ratio (Riechers et al. 2006),
however this must be confirmed with larger samples.

\section{Problems and perspectives}

To tackle galaxy evolution, and the star formation efficiency
as a function of redshift, one of the main problem beyond
sensitivity is the identification of submillimeter sources detected
in continuum  (only 5-10\% are identified). Surveys carried out in 
the far infrared suffer from confusion (Spitzer, ASTRO-F, Herschel). 
  The contribution of AGN to the infrared luminosity has to
be determined.
In the future, the  redshift of the sources could be obtained directly 
from molecular lines ("redshift machine" on LMT, GBT, CCAT, ALMA...). 

In particular,  with a  spatial resolution $<$ 0.1",
there will be no confusion with ALMA. Its sensitivity
will allow the detection of non ULIRGs  and more ``normal'' galaxies, for
example Lyman-Break Galaxies (Steidel et al 1996, Adelberger \& Steidel 2000),
which are observed optically with a density of
150/arcmin$^2$ at  z=2.5-3.5. There will be a complement
to MUSE on the VLT, providing large samples of 
Lyman-$\alpha$  emitters.

\begin{acknowledgements} We wish to thank the IRAM staff
for help with the observations, both at the 30m and Plateau de Bure
interferometer.
IRAM is supported by INSU/CNRS (France), MPG (Germany), and IGN (Spain).
\end{acknowledgements}

\end{document}